\documentclass[12pt]{article}
\usepackage{amssymb}
\usepackage{graphicx}
\begin{document}
\newcommand{\beq}{\begin{equation}}
\newcommand{\eeq}{\end{equation}}
\newcommand{\beqa}{\begin{eqnarray}}
\newcommand{\eeqa}{\end{eqnarray}}
\newcommand{\beqar}{\begin{eqnarray*}}
\newcommand{\eeqar}{\end{eqnarray*}}
\newcommand{\al}{\alpha}
\newcommand{\be}{\beta}
\newcommand{\del}{\delta}
\newcommand{\D}{\Delta}
\newcommand{\eps}{\epsilon}
\newcommand{\ga}{\gamma}
\newcommand{\Ga}{\Gamma}
\newcommand{\ka}{\kappa}
\newcommand{\nn}{\nonumber}
\newcommand{\inn}{\!\cdot\!}
\newcommand{\h}{\eta}
\newcommand{\ii}{\iota}
\newcommand{\kk}{\varphi}
\newcommand\F{{}_3F_2}
\newcommand{\la}{\lambda}
\newcommand{\La}{\Lambda}
\newcommand{\na}{\prt}
\newcommand{\Om}{\Omega}
\newcommand{\om}{\omega}
\newcommand{\p}{\Phi}
\newcommand{\sig}{\sigma}
\renewcommand{\t}{\theta}
\newcommand{\z}{\zeta}
\newcommand{\ssc}{\scriptscriptstyle}
\newcommand{\eg}{{\it e.g.,}\ }
\newcommand{\ie}{{\it i.e.,}\ }
\newcommand{\labell}[1]{\label{#1}} 
\newcommand{\reef}[1]{(\ref{#1})}
\newcommand\prt{\partial}
\newcommand\veps{\varepsilon}
\newcommand{\pol}{\varepsilon}
\newcommand\vp{\varphi}
\newcommand\ls{\ell_s}
\newcommand\cF{{\cal F}}
\newcommand\cA{{\cal A}}
\newcommand\cS{{\cal S}}
\newcommand\cT{{\cal T}}
\newcommand\cV{{\cal V}}
\newcommand\cL{{\cal L}}
\newcommand\cM{{\cal M}}
\newcommand\cN{{\cal N}}
\newcommand\cG{{\cal G}}
\newcommand\cK{{\cal K}}
\newcommand\cH{{\cal H}}
\newcommand\cI{{\cal I}}
\newcommand\cJ{{\cal J}}
\newcommand\cl{{\iota}}
\newcommand\cP{{\cal P}}
\newcommand\cQ{{\cal Q}}
\newcommand\cg{{\tilde {{\cal G}}}}
\newcommand\cR{{\cal R}}
\newcommand\cB{{\cal B}}
\newcommand\cO{{\cal O}}
\newcommand\tcO{{\tilde {{\cal O}}}}
\newcommand\bz{\bar{z}}
\newcommand\bb{\bar{b}}
\newcommand\ba{\bar{a}}
\newcommand\bg{\bar{g}}
\newcommand\bc{\bar{c}}
\newcommand\bw{\bar{w}}
\newcommand\bX{\bar{X}}
\newcommand\bK{\bar{K}}
\newcommand\bA{\bar{A}}
\newcommand\bH{\bar{H}}
\newcommand\bF{\bar{F}}
\newcommand\bxi{\bar{\xi}}
\newcommand\bphi{\bar{\phi}}
\newcommand\bpsi{\bar{\psi}}
\newcommand\bprt{\bar{\prt}}
\newcommand\bet{\bar{\eta}}
\newcommand\btau{\bar{\tau}}
\newcommand\hF{\hat{F}}
\newcommand\hA{\hat{A}}
\newcommand\hT{\hat{T}}
\newcommand\htau{\hat{\tau}}
\newcommand\hD{\hat{D}}
\newcommand\hf{\hat{f}}
\newcommand\hK{\hat{K}}
\newcommand\hg{\hat{g}}
\newcommand\hp{\hat{\Phi}}
\newcommand\hi{\hat{i}}
\newcommand\ha{\hat{a}}
\newcommand\hb{\hat{b}}
\newcommand\hQ{\hat{Q}}
\newcommand\hP{\hat{\Phi}}
\newcommand\hS{\hat{S}}
\newcommand\hX{\hat{X}}
\newcommand\tL{\tilde{\cal L}}
\newcommand\hL{\hat{\cal L}}
\newcommand\tG{{\tilde G}}
\newcommand\tg{{\tilde g}}
\newcommand\tphi{{\widetilde \Phi}}
\newcommand\tPhi{{\widetilde \Phi}}
\newcommand\te{{\tilde e}}
\newcommand\tk{{\tilde k}}
\newcommand\tf{{\tilde f}}
\newcommand\tH{{\tilde H}}
\newcommand\ta{{\tilde a}}
\newcommand\tb{{\tilde b}}
\newcommand\tc{{\tilde c}}
\newcommand\td{{\tilde d}}
\newcommand\tm{{\tilde m}}
\newcommand\tmu{{\tilde \mu}}
\newcommand\tnu{{\tilde \nu}}
\newcommand\talpha{{\tilde \alpha}}
\newcommand\tbeta{{\tilde \beta}}
\newcommand\trho{{\tilde \rho}}
 \newcommand\tR{{\tilde R}}
\newcommand\teta{{\tilde \eta}}
\newcommand\tF{{\widetilde F}}
\newcommand\tK{{\tilde K}}
\newcommand\tE{{\widetilde E}}
\newcommand\tpsi{{\tilde \psi}}
\newcommand\tX{{\widetilde X}}
\newcommand\tD{{\widetilde D}}
\newcommand\tO{{\widetilde O}}
\newcommand\tS{{\tilde S}}
\newcommand\tB{{\tilde B}}
\newcommand\tA{{\widetilde A}}
\newcommand\tT{{\widetilde T}}
\newcommand\tC{{\widetilde C}}
\newcommand\tV{{\widetilde V}}
\newcommand\thF{{\widetilde {\hat {F}}}}
\newcommand\Tr{{\rm Tr}}
\newcommand\tr{{\rm tr}}
\newcommand\STr{{\rm STr}}
\newcommand\hR{\hat{R}}
\newcommand\M[2]{M^{#1}{}_{#2}}
\newcommand\MZ{\mathbb{Z}}
\newcommand\MR{\mathbb{R}}
\newcommand\bS{\textbf{ S}}
\newcommand\bI{\textbf{ I}}
\newcommand\bJ{\textbf{ J}}

\begin{titlepage}
\begin{center}

\vskip 0.5 cm
{\LARGE \bf  
Dimensional reduction of the M-theory\\  \vskip 0.25 cm  Chern-Simons term at order $\ell_p^6$}\\
\vskip 1.25 cm
 Mohammad R. Garousi \footnote{garousi@um.ac.ir}

\vskip 1 cm
{{\it Department of Physics, Faculty of Science, Ferdowsi University of Mashhad\\}{\it P.O. Box 1436, Mashhad, Iran}\\}
\vskip .1 cm
 \end{center}

\begin{abstract}

    The dimensional reduction of M-theory couplings at order $\ell_p^6$ is known to produce one-loop $\alpha'^3$ corrections in type IIA string theory. In this paper, we perform the Kaluza-Klein reduction of the M-theory Chern-Simons coupling $t_8\epsilon_{11} A R^4$ at this order. By meticulously accounting for non-gauge-invariant total derivative terms, we derive the complete set of corresponding one-loop, gauge-invariant couplings in the type IIA effective action. Our results not only reproduce the standard Chern-Simons term $t_8\epsilon_{10} B R^4$—which is gauge invariant up to total derivatives—but also unveil a new set of gauge-invariant couplings involving RR and NS-NS field strengths.

    To validate our findings, we test their consistency under string dualities. We dimensionally reduce the derived type IIA couplings on a K3 manifold and show that the resulting one-loop $\alpha'$ corrections in six dimensions transform under S-duality into the dimensional reduction of the tree-level heterotic string Chern-Simons coupling $H_{\mu\nu\alpha} \Omega^{\mu\nu\alpha}$ on $T^4$. This non-trivial agreement provides strong evidence for the correctness of both the M-theory Chern-Simons term and its reduction to type IIA.

\end{abstract}

\end{titlepage}

\section{Introduction}

M-theory, the hypothesized non-perturbative unification of the five superstring theories \cite{Schwarz:1982jn,Gross:1986iv,Schwarz:1996bh}, is a parity-invariant theory \cite{Duff:1986hr} that is described at low energies by eleven-dimensional supergravity \cite{Huq:1983im,Becker:2007zj}. The leading-order action is well-known, consisting of the Einstein-Hilbert term, a kinetic term for the odd-parity 3-form gauge field $A^{(3)}$, and a crucial cubic Chern-Simons term, $\int A^{(3)} \wedge F^{(4)} \wedge F^{(4)}$. However, this two-derivative action is merely the first term in an infinite series of higher-derivative corrections. These $\ell_p$-suppressed corrections are essential for capturing M-theory's full non-perturbative dynamics and are intimately connected to quantum effects and anomaly cancellation.

The next-order corrections occur at the eight-derivative ($\ell_p^6$) level. The effective action at this order receives contributions from several terms, including the well-known  $(t_8 t_8 - \frac{1}{4!} \epsilon_{11} \epsilon_{11}) R^4$ structures \cite{Vafa:1995fj,Green:1997di,Russo:1997mk,Peeters:2000qj}. A particularly important component is the topological Chern-Simons coupling, schematically denoted $t_8 \epsilon_{11} A  R^4$ \cite{Duff:1995wd,Peeters:2000qj,Hyakutake:2006aq}. Like its lower-order counterpart, this term is gauge-invariant only up to a total derivative and plays a vital role in the generalized Green-Schwarz mechanism for anomaly cancellation in M-theory \cite{Duff:1995wd}. The precise form of this coupling was established through two complementary approaches: consistency conditions of M2-brane/M5-brane duality \cite{Duff:1995wd} and the requirement of local supersymmetry for the purely gravitational  $(t_8 t_8 - \frac{1}{4!} \epsilon_{11} \epsilon_{11}) R^4$ terms \cite{Peeters:2000qj,Hyakutake:2006aq}.

The profound dualities connecting M-theory to string theories provide a powerful tool for exploring these higher-derivative terms. The circular reduction of M-theory yields type IIA string theory \cite{Schwarz:1996bh}, and further compactification on a K3 surface is dual to the heterotic string on $T^4$ \cite{Hull:1994ys,Vafa:1995fj}. The process of Kaluza-Klein (KK) reduction allows us to derive the structure of the one-loop type IIA effective action at order $\alpha'^3$ from this M-theory starting point. The reduction of the pure gravity sector, which produces couplings in the metric-dilaton-RR one-form sector, has been studied recently \cite{Aggarwal:2025lxf,Liu:2025uqu,Garousi:2025wfk} and has been shown to be consistent with string amplitude calculations and heterotic duality.

While it is known that the M-theory Chern-Simons term $t_8\epsilon_{11} A  R^4$ reduces to its type IIA counterpart $t_8\epsilon_{10} B  R^4$ \cite{Duff:1995wd}, a complete classification of all the resulting effective couplings has been absent. In this work, we perform a detailed KK reduction of this term. The direct reduction yields 1,173 non-gauge-invariant couplings in addition to the expected $t_8\epsilon_{10} B  R^4$ term. We demonstrate that these extra couplings can be assembled into a gauge-invariant form by systematically accounting for non-gauge-invariant total derivatives in ten dimensions. Constructing a complete basis of 288 gauge-invariant terms—comprising the Riemann tensor, dilaton derivatives, and the field strengths $H^{(3)}$, $F^{(2)}$, and $\bar{F}^{(4)}$ (where $H^{(3)}$ and $\bar{F}^{(4)}$ appear linearly)—we find that only 91 terms in this basis have non-zero coefficients. The new couplings consist of 11 terms involving the NS-NS field strength $H^{(3)}$ and 80 terms involving the RR field strength $\bar{F}^{(4)}$.

We observe that symmetry considerations do not allow M-theory couplings with a linear dependence on the field strength $F^{(4)}$. Therefore, the reduction of the Chern-Simons couplings yields the only couplings in type IIA theory that are linear in $H^{(3)}$ and $\bar{F}^{(4)}$. This unique property makes them ideal for study under string duality.
We verify that our resulting one-loop couplings in type IIA theory satisfy the constraints imposed by six-dimensional duality with the heterotic string. Upon compactification on K3, these couplings must match the tree-level four-derivative couplings of the heterotic string on $T^4$. We show that this matching occurs exactly, providing a robust consistency check on our results.

This paper is structured as follows. Section 2 details the KK reduction of the M-theory Chern-Simons term at order $\ell_p^6$ and the subsequent construction of the one-loop, gauge-invariant type IIA effective action at order $\alpha'^3$. In Section 3, we explore K3 compactification and S-duality: Subsection 3.1 analyzes the reduction of the resulting type IIA couplings, while Subsection 3.2 studies the corresponding tree-level couplings in the heterotic theory and their reduction on $T^4$. We demonstrate their precise agreement under duality. Our conclusions are presented in Section 4. All calculations were performed using the ``xAct'' package \cite{Nutma:2013zea}.

\section{Reduction of M-theory on  circle}

The well-known conjectured duality between M-theory on a circle and type IIA string theory is reflected in their effective actions: the KK reduction of 11-dimensional supergravity yields 10-dimensional type IIA supergravity \cite{Huq:1983im,Becker:2007zj}, and M-theory's higher-derivative corrections produce quantum corrections in type IIA string theory. The non-zero KK modes are expected to generate stringy corrections to the effective action at each genus level \cite{Green:1997as}, which are not the focus of this work. 

To establish our conventions, we begin by outlining the derivation of the bosonic couplings in 10-dimensional type IIA supergravity from the circular reduction of 11-dimensional supergravity \cite{Huq:1983im,Becker:2007zj}. The bosonic sector of the 11-dimensional theory consists of the following parity-invariant couplings \cite{Becker:2007zj}:
\beqa
\bS_0&=&-\frac{2}{\kappa_{11}^2}\Big[\int d^{11}x\sqrt{-g}(R-\frac{1}{2.4!}F_{abcd}F^{abcd})-\frac{1}{6}\int A^{(3)}\wedge F^{(4)}\wedge F^{(4)}\Big],\labell{11sugra}
\eeqa
where $\kappa_{11}^2=\frac{1}{\pi}(2\pi\ell_p)^9$.
The dimensional reduction of this action on a circle of radius \( R_{11} = g_s^{2/3} \ell_p \) proceeds via the standard KK ansatz for the metric and three-form:
\beqa
g_{ab}=e^{-2\Phi/3}\left(\matrix{G_{\mu\nu}+e^{2\Phi}C_\mu C_\nu & e^{2\Phi}C_\mu &\cr e^{2\Phi}C_\nu &e^{2\Phi}&}\right)\,\,;\,\,A_{\mu\nu\alpha}=C_{\mu\nu\alpha}\,\,;\,\,
A_{\mu\nu y}=B_{\mu\nu}\,.\labell{reduc}
\eeqa
Here, \( G_{\mu\nu} \) denotes the 10-dimensional metric, $\Phi$ represents the 10-dimensional dilaton, \( C^{(3)} \) denotes the RR three-form potential, $C^{(1)}$ denotes the RR one-form potential, and \( B^{(2)} \) represents the NS-NS two-form field of type IIA superstring theory. Under the parity transformation along the circle, $C^{(1)}, C^{(3)}$ are odd, while $B^{(2)}$ is even. Using these reduction rules, one finds the bosonic sector of the parity-invariant 10-dimensional type IIA supergravity, which is  \cite{Becker:2007zj}
\beqa
S_0&=&-\frac{2}{\kappa_{10}^2}\Big[\int d^{10}x \sqrt{-G}\,e^{-2\Phi}\left(R+4\nabla_\mu\Phi\nabla^\mu\Phi-
\frac{1}{2}|H|^2-\frac{e^{2\Phi}}{2}|F^{(2)}|^2-\frac{e^{2\Phi}}{2}|\bar{F}^{(4)}|^2\right)\nn\\
&&\qquad\qquad\qquad\qquad\qquad\qquad\qquad\qquad\qquad\qquad\qquad-\frac{1}{2}\int B^{(2)}\wedge F^{(4)}\wedge F^{(4)}\Big].\labell{S0}
\eeqa
Here,  \( F_{\mu\nu} =2 \partial_{[\mu }C_{\nu]}  \),  $H_{\mu\nu\alpha}=3\prt_{[\mu}B_{\nu\alpha]}$, 
$
\bar{F}_{\mu\nu\alpha\beta} = 4\prt_{[\mu}C_{\nu\alpha\beta]} + 4C_{[\mu}H_{\nu\alpha\beta]}
$, and the 10-dimensional gravitational coupling is given by \(\kappa_{10}^2 = \kappa^2 g_s^2 = \frac{1}{\pi}(2\pi\ell_s)^8 g_s^2\), where \(\ell_s = \sqrt{\alpha'}\) defines the string length.  Note that the 10-dimensional Levi-Civita tensor in the last term has no component along the circle direction; it is therefore even under the parity transformation. Hence, the action is invariant. Also note that each RR field is accompanied by a dilaton factor of \(e^{\Phi}\). With the overall dilaton factor of the action being \(e^{-2\Phi}\), this confirms it corresponds to the sphere-level effective action.

The first non-trivial higher-derivative correction to 11-dimensional supergravity appears at eight-derivative order (corresponding to \(\mathcal{O}(\ell_p^6)\)). While the complete structure of bosonic couplings at this order remains unknown, the purely gravitational sector and Chern-Simons term have been determined and takes the form (see \eg  \cite{Hyakutake:2007vc}):
\beqa
\bS_6&=&-\frac{2}{\kappa_{11}^2}\frac{\pi^2\ell_p^6}{2^{11}.3^2}\int d^{11}x\sqrt{-g}\Big[(t_8t_8-\frac{1}{4!}\epsilon_{11}\epsilon_{11})R^4-\frac{1}{6}t_8\epsilon_{11}AR^4+\cdots\Big],\labell{ttee}
\eeqa
where the ellipsis denotes additional couplings involving higher orders of  the 3-form field \(A_{abc}\). The reduction of the pure gravitational part has been studied in \cite{Aggarwal:2025lxf,Garousi:2025wfk}. Here we are interested in reduction of the Chern-Simons term. 
The tensor \(t_8\) is defined as \cite{Schwarz:1982jn}:
\beqa
t_8^{abcdefgh}&=&-2 g^{a f } g^{b e } g^{g d } 
g^{c h } + 8 g^{a d } g^{b e } 
g^{g f } g^{c h } + 8 g^{a h } g^{b 
e } g^{g d } g^{c f } \nn\\&&+ 8 
g^{a h } g^{b c } g^{g f } g^{d 
e } - 2 g^{a h } g^{b g } g^{c 
f } g^{d e } - 2 g^{a d } 
g^{b c } g^{g f } g^{e h }\,,
\eeqa
 and $\epsilon_{11}$ denotes the odd-parity Levi-Civita tensor in eleven dimensions. Using these tensors, the Chern-Simons term  can be expressed as
\beqa
\bS^{CS}_6&=&-\frac{2}{\kappa_{11}^2}\frac{\pi^2\ell_p^6}{2^{11}.3^2}\int d^{11}x\sqrt{-g}\,\epsilon_{11}^{abcdefghijk}A_{abc} \Big[ R_{pnhi} R^{pn}{}_{jk} R_{qmde} 
R^{qm}{}_{fg} \nn\\&&\qquad\qquad\qquad\qquad\qquad\qquad\qquad\qquad\quad- 4 R^{m}{}_{nhi} 
R^{pn}{}_{jk} R_{qmde} R^{q}{}_{pfg}\Big].\labell{ttee1}
\eeqa
The Levi-Civita tensor $\epsilon_{11}$ is related to the Levi-Civita symbol $\epsilon'_{11}$ as $\sqrt{-g}\,\epsilon_{11}=\epsilon'_{11}$.

The reduction of the three-form is given in \reef{reduc}, and the reduction of the Levi-Civita symbol is trivial, as
\beqa
\sqrt{-g} \,\epsilon_{11}^{y\alpha \beta \gamma \mu \nu \kappa
\lambda \theta \delta \sigma }=\sqrt{-G} \,\epsilon_{10}^{\alpha \beta \gamma \mu \nu \kappa
\lambda \theta \delta \sigma },
\eeqa
where $\epsilon_{10}$ denotes the Levi-Civita tensor in ten dimensions. The reduction of the Riemann curvature tensor can also be calculated using the reduction of the metric in \reef{reduc}. However, it includes many terms that depend on both the RR one-form potential and its field strength. The reduction of this tensor, when all indices of the 11-dimensional object are 10-dimensional indices, is
\beqa
R_{\alpha \beta \gamma \delta }&\!\!\!\!\!=\!\!\!\!\!&e^{-2 \Phi/3} R_{\alpha \beta \gamma \delta } 
+ \frac{1}{4} e^{4 \Phi/3} F_{\alpha \delta } F_{\beta \gamma 
} -  \frac{1}{4} e^{4 \Phi/3} F_{\alpha \gamma } F_{\beta 
\delta } -  \frac{1}{2} e^{4 \Phi/3} F_{\alpha \beta } 
F_{\gamma \delta } + \frac{1}{4} e^{10 \Phi/3} F_{\beta 
}{}^{\mu } F_{\delta \mu } C_{\alpha } C_{\gamma }\nn\\&& -  
\frac{1}{4} e^{10 \Phi/3} F_{\alpha }{}^{\mu } F_{\delta \mu } 
C_{\beta } C_{\gamma } -  \frac{1}{4} e^{10 \Phi/3} F_{\beta 
}{}^{\mu } F_{\gamma \mu } C_{\alpha } C_{\delta } + 
\frac{1}{4} e^{10 \Phi/3} F_{\alpha }{}^{\mu } F_{\gamma \mu } 
C_{\beta } C_{\delta } -  \frac{1}{2} e^{4 \Phi/3} C_{\gamma 
} \nabla_{\alpha }F_{\beta \delta }\nn\\&& -  \frac{1}{2} e^{4
\Phi/3} C_{\beta } \nabla_{\alpha }F_{\gamma \delta } -  e^{4
\Phi/3} F_{\gamma \delta } C_{\beta } \nabla_{\alpha }\Phi -  
\frac{1}{2} e^{4 \Phi/3} F_{\beta \delta } C_{\gamma } 
\nabla_{\alpha }\Phi + \frac{1}{2} e^{4 \Phi/3} F_{\beta 
\gamma } C_{\delta } \nabla_{\alpha }\Phi \nn\\&&+ \frac{1}{2} 
e^{4 \Phi/3} C_{\gamma } \nabla_{\beta }F_{\alpha \delta } + 
e^{4 \Phi/3} F_{\gamma \delta } C_{\alpha } \nabla_{\beta 
}\Phi + \frac{1}{2} e^{4 \Phi/3} F_{\alpha \delta } C_{\gamma 
} \nabla_{\beta }\Phi -  \frac{1}{2} e^{4 \Phi/3} F_{\alpha 
\gamma } C_{\delta } \nabla_{\beta }\Phi \nn\\&&-  \frac{1}{2} 
e^{4 \Phi/3} C_{\delta } \nabla_{\gamma }F_{\alpha \beta } + 
\frac{1}{2} e^{4 \Phi/3} C_{\alpha } \nabla_{\gamma }F_{\beta 
\delta } + \frac{1}{2} e^{4 \Phi/3} F_{\beta \delta } 
C_{\alpha } \nabla_{\gamma }\Phi -  \frac{1}{2} e^{4 \Phi/3} 
F_{\alpha \delta } C_{\beta } \nabla_{\gamma }\Phi \nn\\&&-  e^{4 
\Phi/3} F_{\alpha \beta } C_{\delta } \nabla_{\gamma }\Phi + 
\frac{1}{9} e^{-2 \Phi/3} G_{\beta \delta } \nabla_{\alpha }\Phi \nabla_{\gamma }
\Phi -  \frac{8}{9} e^{4 \Phi/3} C_{\beta } 
C_{\delta } \nabla_{\alpha }\Phi \nabla_{\gamma }\Phi \nn\\&&-  
\frac{1}{9} e^{-2 \Phi/3} G_{\alpha \delta } \nabla_{\beta }\Phi \nabla_{\gamma }
\Phi+ \frac{8}{9} e^{4 \Phi/3} C_{\alpha } 
C_{\delta } \nabla_{\beta }\Phi \nabla_{\gamma }\Phi + 
\frac{1}{3} e^{-2 \Phi/3} G_{\beta \delta } \nabla_{\gamma }\nabla_{\alpha 
}\Phi \nn\\&&-  \frac{2}{3} e^{4 \Phi/3} C_{\beta } 
C_{\delta } \nabla_{\gamma }\nabla_{\alpha }\Phi -  \frac{1}{3} e^{-2 
\Phi/3}G_{
\alpha \delta } \nabla_{\gamma }\nabla_{\beta }\Phi + \frac{2}{3} e^{4 \Phi/3} C_{\alpha } C_{\delta } 
\nabla_{\gamma }\nabla_{\beta }\Phi -  \frac{1}{2} e^{4
\Phi/3} C_{\alpha } \nabla_{\delta }F_{\beta \gamma }\nn\\&&-  
\frac{1}{2} e^{4 \Phi/3} F_{\beta \gamma } C_{\alpha } 
\nabla_{\delta }\Phi + \frac{1}{2} e^{4 \Phi/3} F_{\alpha 
\gamma } C_{\beta } \nabla_{\delta }\Phi + e^{4 \Phi/3} 
F_{\alpha \beta } C_{\gamma } \nabla_{\delta }\Phi -  
\frac{1}{9} e^{-2 \Phi/3} G_{\beta \gamma } \nabla_{\alpha }\Phi \nabla_{\delta }
\Phi\nn\\&& + \frac{8}{9} e^{4 \Phi/3} C_{\beta } 
C_{\gamma } \nabla_{\alpha }\Phi \nabla_{\delta }\Phi + 
\frac{1}{9} e^{-2 \Phi/3} G_{\alpha \gamma } \nabla_{\beta }\Phi \nabla_{\delta }
\Phi -  \frac{8}{9} e^{4 \Phi/3} C_{\alpha } 
C_{\gamma } \nabla_{\beta }\Phi \nabla_{\delta }\Phi \nn\\&&-  
\frac{1}{3} e^{-2 \Phi/3} G_{\beta \gamma } \nabla_{\delta }\nabla_{\alpha 
}\Phi + \frac{2}{3} e^{4 \Phi/3} C_{\beta } 
C_{\gamma } \nabla_{\delta }\nabla_{\alpha }\Phi + 
\frac{1}{3} e^{-2 \Phi/3} G_{\alpha \gamma } \nabla_{\delta }\nabla_{\beta 
}\Phi \labell{R4}\\&&-  \frac{2}{3} e^{4 \Phi/3} C_{\alpha } 
C_{\gamma } \nabla_{\delta }\nabla_{\beta }\Phi -  
\frac{1}{6} e^{4 \Phi/3} G_{\beta \delta } F_{\gamma \mu } 
C_{\alpha } \nabla^{\mu }\Phi + \frac{1}{6} e^{4 \Phi/3} 
G_{\beta \gamma } F_{\delta \mu } C_{\alpha } \nabla^{\mu 
}\Phi \nn\\&&+ \frac{1}{6} e^{4 \Phi/3} G_{\alpha \delta } F_{\gamma 
\mu } C_{\beta } \nabla^{\mu }\Phi -  \frac{1}{6} e^{4
\Phi/3} G_{\alpha \gamma } F_{\delta \mu } C_{\beta } 
\nabla^{\mu }\Phi -  \frac{1}{6} e^{4 \Phi/3} G_{\beta \delta 
} F_{\alpha \mu } C_{\gamma } \nabla^{\mu }\Phi  \nn\\&&+ 
\frac{1}{6} e^{4 \Phi/3} G_{\alpha \delta } F_{\beta \mu } 
C_{\gamma } \nabla^{\mu }\Phi+ \frac{1}{6} e^{4 \Phi/3} 
G_{\beta \gamma } F_{\alpha \mu } C_{\delta } \nabla^{\mu 
}\Phi -  \frac{1}{6} e^{4 \Phi/3} G_{\alpha \gamma } F_{\beta 
\mu } C_{\delta } \nabla^{\mu }\Phi  \nn\\&&+ \frac{1}{9} 
e^{-2 \Phi/3} G_{\alpha \delta 
} G_{\beta \gamma } \nabla_{\mu }\Phi \nabla^{\mu }\Phi -  \frac{1}{9} e^{-2 \Phi/3} G_{\alpha \gamma } G_{\beta \delta } 
\nabla_{\mu }\Phi \nabla^{\mu }\Phi + 
\frac{2}{9} e^{4 \Phi/3} G_{\beta \delta } C_{\alpha } 
C_{\gamma } \nabla_{\mu }\Phi \nabla^{\mu }\Phi \nn\\&& -  
\frac{2}{9} e^{4 \Phi/3} G_{\alpha \delta } C_{\beta } 
C_{\gamma } \nabla_{\mu }\Phi \nabla^{\mu }\Phi -  
\frac{2}{9} e^{4 \Phi/3} G_{\beta \gamma } C_{\alpha } 
C_{\delta } \nabla_{\mu }\Phi \nabla^{\mu }\Phi + 
\frac{2}{9} e^{4 \Phi/3} G_{\alpha \gamma } C_{\beta } 
C_{\delta } \nabla_{\mu }\Phi \nabla^{\mu }\Phi .\nn
\eeqa
Note that we have used the same symbol for both the 11-dimensional and 10-dimensional Riemann curvature. The reduction of the Riemann curvature, when three indices are 10-dimensional indices and one of them is the Killing index $y$, becomes
\beqa
R_{\alpha \beta \gamma y }&\!\!\!=\!\!\!&- \frac{1}{4} e^{10 \Phi/3} F_{\beta }{}^{\delta } F_{\gamma 
\delta } C_{\alpha } + \frac{1}{4} e^{10 \Phi/3} F_{\alpha 
}{}^{\delta } F_{\gamma \delta } C_{\beta } + \frac{1}{2} 
e^{4 \Phi/3} F_{\beta \gamma } \nabla_{\alpha }\Phi -  
\frac{1}{2} e^{4 \Phi/3} F_{\alpha \gamma } \nabla_{\beta 
}\Phi \nn\\&&+ \frac{1}{2} e^{4 \Phi/3} \nabla_{\gamma }F_{\beta 
\alpha } -  e^{4 \Phi/3} F_{\alpha \beta } \nabla_{\gamma 
}\Phi -  \frac{8}{9} e^{4 \Phi/3} C_{\beta } \nabla_{\alpha }
\Phi \nabla_{\gamma }\Phi + \frac{8}{9} e^{4 \Phi/3} 
C_{\alpha } \nabla_{\beta }\Phi \nabla_{\gamma }\Phi\nn\\&& -  
\frac{2}{3} e^{4 \Phi/3} C_{\beta } \nabla_{\gamma 
}\nabla_{\alpha }\Phi + \frac{2}{3} e^{4 \Phi/3} C_{\alpha } 
\nabla_{\gamma }\nabla_{\beta }\Phi + \frac{1}{6} e^{4
\Phi/3} G_{\beta \gamma } F_{\alpha \delta } \nabla^{\delta 
}\Phi -  \frac{1}{6} e^{4 \Phi/3} G_{\alpha \gamma } F_{\beta 
\delta } \nabla^{\delta }\Phi \nn\\&&-  \frac{2}{9} e^{4 \Phi/3} 
G_{\beta \gamma } C_{\alpha } \nabla_{\delta }\Phi 
\nabla^{\delta }\Phi + \frac{2}{9} e^{4 \Phi/3} G_{\alpha 
\gamma } C_{\beta } \nabla_{\delta }\Phi \nabla^{\delta 
}\Phi.
\eeqa
The reduction of the Riemann curvature, when two indices are 10-dimensional indices and two of them are the Killing index $y$, becomes
\beqa
R_{\alpha y \gamma y }&\!\!\!=\!\!\!&\frac{1}{4} e^{10 \Phi/3} F_{\alpha }{}^{\beta } F_{\gamma 
\beta } + \frac{2}{9} e^{4 \Phi/3} G_{\alpha \gamma } \nabla_{
\beta }\Phi \nabla^{\beta }\Phi -  \frac{8}{9} e^{4 \Phi/3} 
\nabla_{\alpha }\Phi \nabla_{\gamma }\Phi -  \frac{2}{3} 
e^{4 \Phi/3} \nabla_{\gamma }\nabla_{\alpha }\Phi ,
\eeqa
which does not contain the RR one-form potential.  If one considers the above reductions, then the reduction of the action \reef{ttee1} can, in principle, be calculated. However, given that the reduction of the Riemann curvature with all indices in the 10-dimensional space has 56 terms, the reduction of four Riemann curvature tensors produces too many terms to be analyzed with an ordinary computer. Nevertheless, by using the fact that the final result must be gauge invariant up to total derivative terms, one can remove the terms that would cancel each other in the end, before performing the explicit calculation.

Similar complications arise when one attempts to calculate the KK reduction of the pure gravity sector in \reef{ttee}. In this case, the final result must be gauge invariant without relying on any total derivative terms. Consequently, one can simplify the calculation significantly by initially setting all RR one-form potentials —which appear in the curvature reduction —to zero. This approach is justified because any such terms would ultimately need to cancel in the final ten-dimensional action after a lengthy computation. By implementing this simplification, the calculation becomes tractable and can be performed using an ordinary computer \cite{Garousi:2025wfk}.

In the reduction of the Chern-Simons term, however, the result must be gauge invariant only after incorporating appropriate non-gauge invariant total derivative terms. Hence, the RR one-form potentials cannot be removed from the Riemann curvature reduction. Using the fact that the final result—after incorporating non-gauge invariant total derivative terms—must be gauge invariant and be written linearly in terms of the NS-NS field strength $H^{(3)}$ and the RR field strength $\bar{F}^{(4)}=dC^{(3)}+C^{(1)}\wedge H^{(3)}$, the RR one-form potential must appear only linearly in the reduction of the coupling \reef{ttee1}. This allows us, during the reduction process, to remove all terms that have two or more factors of the RR one-form potential, which greatly simplifies the calculation.
 
Through the application of the dimensional reduction described above, we arrive at the following couplings in the string frame of type IIA theory:
\beqa
S^{CS}_6&=&-\frac{2}{\kappa^2}\frac{\pi^2\ell_s^6}{2^{11}.3^2}\int d^{10}x\sqrt{-G} \,\epsilon_{10}^{\alpha \beta \gamma \mu \nu \kappa 
\lambda \theta \delta \sigma }\Big[3 B_{\alpha \beta } R_{\gamma \mu 
}{}^{\eta \epsilon } R_{\lambda \theta \rho \omega } 
R_{\nu \kappa \eta \epsilon } R_{
\delta\sigma }{}^{\rho \omega }\labell{reduce}\\&&\qquad\qquad\qquad\qquad\qquad\qquad\qquad\quad- 12 B_{\alpha \beta }  
R_{\gamma \mu }{}^{\eta \epsilon } R_{\lambda 
\theta \omega \epsilon } R_{\nu \kappa \rho \eta } R_{\delta\sigma }{}^{\rho \omega }+\cL_{\alpha \beta \gamma \mu \nu \kappa 
\lambda \theta \delta \sigma}\Big].\nn
\eeqa
 The absence of an overall dilaton factor indicates that the reduced action $S_6^{CS}$ corresponds to the torus-level effective action of type IIA theory.  $\mathcal{L}_{\alpha \beta \gamma \mu \nu \kappa \lambda \theta \delta \sigma}$ represents 1,173 couplings involving the Riemann tensor, along with first and second derivatives of the dilaton and the RR one-form field strength. They also include many terms that involve the RR three-form and one-form potentials as well as the NS-NS two-form potential. While the RR three-form and the NS-NS two-form potential naturally appear in the reduction only linearly, we removed second and higher orders of the RR one-form potential, as we argued in favor of this approach in the previous paragraph. To ensure that this assumption is legitimate, we performed a similar calculation in six dimensions which involved two Riemann tensors, and found that all nonlinear one-form potential terms canceled after using the Bianchi identities:
\beqa
\nabla_{[\mu}F_{\nu\alpha]}\,=\,0\,,&& R_{\mu[\nu\alpha\beta]}\,=\,0.\label{Bian}
\eeqa
Each RR field in \reef{reduce} appears with a dilaton factor $e^{\Phi}$, as expected.

Terms in the KK reduction of \reef{ttee1} that contain the RR one-form potential appear in conjunction with the NS-NS two-form potential. In fact, terms where the RR one-form and three-form potentials appear together are canceled upon applying the aforementioned Bianchi identities. The reduction \reef{reduce} contains no such terms.
Furthermore, in all these terms, the RR one-form potential contracts exclusively with the Levi-Civita tensor. This property allows the terms to satisfy the RR gauge invariance condition after appropriate non-gauge-invariant total derivative terms are included.
While the reduction of \reef{ttee1} produces many terms where the RR one-form potential does not contract with the Levi-Civita tensor, all such terms cancel when the Bianchi identities are applied. Crucially, if the RR one-form potential were to contract with another tensor, the resulting terms would violate RR gauge invariance, even after the addition of non-gauge-invariant total derivative terms. The reduction \reef{reduce} contains no terms of this kind.

Since the original Chern-Simons term is invariant under the three-form gauge transformation up to total derivative terms, one expects the reduced terms in \reef{reduce} to satisfy the RR and NS-NS gauge transformations up to some total derivative terms. In other words, if one adds some appropriate non-gauge invariant total derivative terms to the 10-dimensional couplings in \reef{reduce}, the result should be written in terms of the RR and NS-NS field strengths, which are gauge invariant. This is except for the two $BR^4$ terms in \reef{reduce}, which cannot be written in terms of the Riemann curvature and the NS-NS field strength. However, all other 1,173 terms can be written in terms of the Riemann curvature and the RR and NS-NS field strengths.

Having found the reduction of the Chern-Simons coupling in \reef{reduce}, one must now add some non-gauge-invariant total derivative terms to rewrite the result in a gauge-invariant form. We consider the following total derivative terms:
\beqa
\cJ=-\frac{2}{\kappa^2}\frac{\pi^2\ell_s^6}{2^{11}.3^2}\int d^{10}x\sqrt{-G}\,\nabla_\mu\cI^\mu,
\eeqa 
where the vector $\mathcal{I}^\mu$ is constructed from all seven-derivative contractions of the Riemann tensor, the first and second derivatives of the dilaton, $F_{\mu\nu}$ and its first derivative, and the linear terms of $C^{(3)}$, $B^{(2)}$, $C^{(1)}$, and $B^{(2)}C^{(1)}$. The potentials must contract with the 10-dimensional Levi-Civita tensor. All vectors constructed in this way contain arbitrary coefficients.

If one adds the above total derivative terms to the reduction \reef{reduce}, then by choosing the appropriate values for the arbitrary coefficients of the total derivative terms, one may be able to write the non-gauge-invariant couplings \reef{reduce} in a gauge-invariant form. However, to perform this step, it is appropriate to find a basis for gauge-invariant couplings in order to write the resulting gauge-invariant couplings in terms of this basis. To find this basis, we consider
 \beqa
S^{Basis}_6&=&-\frac{2}{\kappa^2}\frac{\pi^2\ell_s^6}{2^{11}.3^2}\int d^{10}x\sqrt{-G}\cL^{Basis}\,,\labell{ttee2}
\eeqa 
where $\mathcal{L}^{Basis}$ contains all eight-derivative contractions of the Riemann curvature tensor, the first and second derivatives of the dilaton, $F_{\mu\nu}$ and its first derivative, and linear terms of the field strengths $H^{(3)}$ and $\bar{F}^{(4)}$ that contract with the 10-dimensional Levi-Civita tensor. We remove terms that involve the Ricci tensor and Ricci scalar, as well as terms involving $\nabla_\mu F^{\mu\nu}$ and $\nabla_\mu\nabla^\mu\Phi$. Removing also the terms that are related by the Bianchi identities \reef{Bian}, one finds the basis consists of 288 couplings with coefficients $a_1,\cdots, a_{288}$. Note that we did not remove terms in the basis which are related by total derivative terms. This means that not all 288 terms in the basis are independent. If one removes all total derivative terms from the basis, one would find 249 independent terms. However, for reasons that we explain in the next paragraph, we prefer to consider the basis with 288 terms.

We then equate the above basis with the reduction in \reef{reduce}, to which the non-gauge-invariant total derivative terms have also been added, \ie,
\beqa
S_6^{CS}+\cJ&=&S^{Basis}_6.
\eeqa
To solve the above equation, we move to the local frame and express the field strengths in terms of potentials. This approach ensures the Bianchi identities are satisfied \cite{Garousi:2019cdn}. We then derive an algebraic equation involving the fixed numbers in the reduction \reef{reduce}, the coefficients $a_1,\cdots, a_{288}$, and the parameters from the total derivative terms. This equation yields 249 relations among the 288 parameters in the basis.
If we had considered a basis with only 249 couplings, this equation would have fixed all 249 parameters, resulting in 171 non-zero terms. However, with the 288-term basis, the equation allows us to solve for 39 of the parameters, enabling the final Lagrangian to be written with fewer than 171 couplings.
We fix the remaining parameters by imposing the constraint that the couplings contain no second derivative of the dilaton. This fixes 28 parameters. With 11 parameters still undetermined, we set them to zero. The final result is a Lagrangian with only 93 non-zero couplings. They are 
\beqa
S^{CS}_6&\!\!\!\!\!=\!\!\!\!\!&-\frac{2}{\kappa^2}\frac{\pi^2\ell_s^6}{2^{11}.3^2}\int d^{10}x\sqrt{-G} \,\epsilon_{10}^{\alpha \beta \gamma \mu \nu \kappa 
\lambda \theta \delta \sigma }\Big[B_{\alpha\beta}\cL^B_{ \gamma \mu \nu \kappa 
\lambda \theta \delta \sigma}\!-\!H_{\alpha \delta \sigma}\cL^H_{ \beta \gamma\mu \nu \kappa 
\lambda \theta }\!-\!\bar{F}_{\alpha \theta \delta \sigma}\cL^{\bar{F}}_{\beta \gamma \mu \nu \kappa 
\lambda }\Big]\!,\labell{reduce1}
\eeqa
in which two standard terms of $\cL^B_{\gamma\mu\nu\kappa\lambda\theta\delta\sigma}$ are
\beqa
\cL^B_{ \gamma \mu \nu \kappa 
\lambda \theta \delta \sigma}&=&3  R_{\gamma \mu 
}{}^{\eta \epsilon } R_{\lambda \theta \rho \omega } 
R_{\nu \kappa \eta \epsilon } R_{\delta\sigma 
 }{}^{\rho \omega } - 12  
R_{\gamma \mu }{}^{\eta \epsilon } R_{\lambda 
\theta \omega \epsilon } R_{\nu \kappa \rho \eta } R_{ \delta\sigma }{}^{\rho \omega }.
\eeqa
The Lagrangian $\cL^H_{\beta\gamma\mu\nu\kappa\lambda\theta}$ comprises 11 terms. They are
\beqa
\cL^H_{ \beta \gamma\mu \nu \kappa 
\lambda \theta }&=&-16 e^{4 \Phi} R_{\lambda \theta \tau \omega } F_{\gamma }{}^{\rho } F_{\mu }{}^{\tau } F_{\nu \kappa } F_{\rho }{}^{\omega } \nabla_{\beta }\Phi - 4 e^{6 \Phi} F_{\beta }{}^{\rho } F_{\gamma \mu } F_{\nu \kappa } F_{\rho }{}^{\tau } F_{\tau }{}^{\omega } \nabla_{\theta }F_{\lambda \omega } \nn\\&&- 4 e^{6 \Phi} F_{\beta }{}^{\rho } F_{\gamma }{}^{\tau } F_{\kappa \lambda } F_{\mu \nu } F_{\rho }{}^{\omega } \nabla_{\theta }F_{\tau \omega } + 6 e^{4 \Phi} R_{\lambda \theta \tau \omega } F_{\beta }{}^{\rho } F_{\gamma }{}^{\tau } F_{\mu }{}^{\omega } \nabla_{\kappa }F_{\nu \rho } \nn\\&&+ 2 e^{4 \Phi} R_{\lambda \theta \tau \omega } F_{\beta }{}^{\rho } F_{\gamma \mu } F^{\tau \omega } \nabla_{\kappa }F_{\nu \rho } - 8 e^{4 \Phi} R_{\lambda \theta \tau \omega } F_{\beta }{}^{\rho } F_{\gamma }{}^{\tau } F_{\mu \nu } \nabla_{\kappa }F_{\rho }{}^{\omega }\nn\\&& + 8 e^{4 \Phi} R_{\lambda \theta \tau \omega } F_{\beta \gamma } F_{\mu \nu } F^{\rho \tau } \nabla_{\kappa }F_{\rho }{}^{\omega } - 4 e^{2 \Phi} R_{\lambda \theta \tau \omega } R_{\nu \kappa }{}^{\tau \omega } F_{\beta }{}^{\rho } \nabla_{\mu }F_{\gamma \rho } \nn\\&&+ 8 e^{2 \Phi} R_{\lambda \theta \tau \omega } R_{\nu \kappa \rho }{}^{\omega } F_{\beta }{}^{\rho } \nabla^{\tau }F_{\gamma \mu } + 2 e^{4 \Phi} F_{\beta }{}^{\rho } \nabla_{\theta }F_{\lambda \tau } \nabla_{\mu }F_{\gamma \rho } \nabla^{\tau }F_{\nu \kappa } \nn\\&&+ 12 e^{4 \Phi} R_{\lambda \theta \tau \omega } F_{\beta }{}^{\rho } F_{\gamma \mu } F_{\rho }{}^{\tau } \nabla^{\omega }F_{\nu \kappa },
\eeqa
and $\cL^{\bar{F}}_{\beta\gamma\mu\nu\kappa\lambda}$ contains 80 terms. They are
\beqa
\cL^{\bar{F}}_{\beta \gamma \mu \nu \kappa 
\lambda }&=&- \frac{1}{2} e^{4 \Phi} R_{\kappa \lambda \omega \eta } R_{\mu \nu }{}^{\omega \eta } F_{\beta }{}^{\rho } F_{\gamma }{}^{\tau } F_{\rho \tau } + \frac{3}{2} e^{6 \Phi} R_{\kappa \lambda \omega \eta } F_{\beta }{}^{\rho } F_{\gamma }{}^{\tau } F_{\mu }{}^{\omega } F_{\nu }{}^{\eta } F_{\rho \tau } \nn\\&&- 4 e^{2 \Phi} R_{\beta \gamma \rho }{}^{\omega } R_{\kappa \lambda \omega \eta } R_{\mu \nu \tau }{}^{\eta } F^{\rho \tau } + e^{2 \Phi} R_{\beta \gamma \rho \tau } R_{\kappa \lambda \omega \eta } R_{\mu \nu }{}^{\omega \eta } F^{\rho \tau }\nn\\&& -  \frac{1}{4} e^{4 \Phi} R_{\kappa \lambda \omega \eta } R_{\mu \nu }{}^{\omega \eta } F_{\beta \gamma } F_{\rho \tau } F^{\rho \tau } - 2 e^{6 \Phi} R_{\kappa \lambda \omega \eta } F_{\beta }{}^{\rho } F_{\gamma }{}^{\tau } F_{\mu \nu } F_{\rho }{}^{\omega } F_{\tau }{}^{\eta } \nn\\&&-  e^{6 \Phi} R_{\kappa \lambda \omega \eta } F_{\beta \gamma } F_{\mu \nu } F_{\rho }{}^{\omega } F^{\rho \tau } F_{\tau }{}^{\eta } -  \frac{3}{8} e^{8 \Phi} F_{\beta }{}^{\rho } F_{\gamma }{}^{\tau } F_{\kappa \lambda } F_{\mu }{}^{\omega } F_{\nu }{}^{\eta } F_{\rho \tau } F_{\omega \eta } \nn\\&&+ \frac{1}{2} e^{8 \Phi} F_{\beta }{}^{\rho } F_{\gamma }{}^{\tau } F_{\kappa \lambda } F_{\mu \nu } F_{\rho }{}^{\omega } F_{\tau }{}^{\eta } F_{\omega \eta } + \frac{1}{8} e^{8 \Phi} F_{\beta \gamma } F_{\kappa \lambda } F_{\mu \nu } F_{\rho }{}^{\omega } F^{\rho \tau } F_{\tau }{}^{\eta } F_{\omega \eta } \nn\\&&-  \frac{1}{2} e^{4 \Phi} R_{\kappa \lambda \rho \tau } R_{\mu \nu \omega \eta } F_{\beta }{}^{\rho } F_{\gamma }{}^{\tau } F^{\omega \eta } + \frac{1}{2} e^{6 \Phi} R_{\kappa \lambda \omega \eta } F_{\beta }{}^{\rho } F_{\gamma }{}^{\tau } F_{\mu \nu } F_{\rho \tau } F^{\omega \eta } \nn\\&&-  \frac{1}{2} e^{4 \Phi} R_{\kappa \lambda \omega \eta } R_{\mu \nu \rho \tau } F_{\beta \gamma } F^{\rho \tau } F^{\omega \eta } + 3 e^{4 \Phi} R_{\kappa \lambda \tau \eta } R_{\mu \nu \rho \omega } F_{\beta \gamma } F^{\rho \tau } F^{\omega \eta } \nn\\&&+ \frac{1}{4} e^{6 \Phi} R_{\kappa \lambda \omega \eta } F_{\beta \gamma } F_{\mu \nu } F_{\rho \tau } F^{\rho \tau } F^{\omega \eta } + \frac{1}{4} e^{6 \Phi} R_{\kappa \lambda \rho \tau } F_{\beta }{}^{\rho } F_{\gamma }{}^{\tau } F_{\mu \nu } F_{\omega \eta } F^{\omega \eta } \nn\\&&-  \frac{1}{8} e^{8 \Phi} F_{\beta }{}^{\rho } F_{\gamma }{}^{\tau } F_{\kappa \lambda } F_{\mu \nu } F_{\rho \tau } F_{\omega \eta } F^{\omega \eta } -  \frac{1}{32} e^{8 \Phi} F_{\beta \gamma } F_{\kappa \lambda } F_{\mu \nu } F_{\rho \tau } F^{\rho \tau } F_{\omega \eta } F^{\omega \eta } \nn\\&&+ e^{6 \Phi} F_{\gamma }{}^{\rho } F_{\mu \nu } F_{\tau \omega } F^{\tau \omega } \nabla_{\beta }\Phi \nabla_{\lambda }F_{\kappa \rho } + 6 e^{6 \Phi} F_{\gamma }{}^{\rho } F_{\mu }{}^{\tau } F_{\nu }{}^{\omega } F_{\rho \tau } \nabla_{\beta }\Phi \nabla_{\lambda }F_{\kappa \omega } \nn\\&&- 8 e^{6 \Phi} F_{\gamma }{}^{\rho } F_{\mu \nu } F_{\rho }{}^{\tau } F_{\tau }{}^{\omega } \nabla_{\beta }\Phi \nabla_{\lambda }F_{\kappa \omega } - 12 e^{6 \Phi} F_{\gamma }{}^{\rho } F_{\mu }{}^{\tau } F_{\nu \kappa } F_{\rho }{}^{\omega } \nabla_{\beta }\Phi \nabla_{\lambda }F_{\tau \omega } \nn\\&&- 2 e^{6 \Phi} F_{\beta }{}^{\rho } F_{\gamma }{}^{\tau } F_{\mu \nu } \nabla_{\kappa }F_{\rho }{}^{\omega } \nabla_{\lambda }F_{\tau \omega } - 8 e^{4 \Phi} R_{\kappa \lambda \tau \omega } F_{\gamma }{}^{\rho } F_{\mu }{}^{\tau } \nabla_{\beta }\Phi \nabla_{\nu }F_{\rho }{}^{\omega } \nn\\&&+ 16 e^{4 \Phi} R_{\kappa \lambda \tau \omega } F_{\gamma \mu } F^{\rho \tau } \nabla_{\beta }\Phi \nabla_{\nu }F_{\rho }{}^{\omega } + 4 e^{2 \Phi} R_{\kappa \lambda \tau \omega } R_{\mu \nu }{}^{\tau \omega } F_{\gamma \rho } \nabla_{\beta }\Phi \nabla^{\rho }\Phi \nn\\&&- 16 e^{2 \Phi} R_{\kappa \lambda \tau \omega } R_{\mu \nu \rho }{}^{\omega } F_{\gamma }{}^{\tau } \nabla_{\beta }\Phi \nabla^{\rho }\Phi - 12 e^{4 \Phi} R_{\kappa \lambda \tau \omega } F_{\gamma \rho } F_{\mu }{}^{\tau } F_{\nu }{}^{\omega } \nabla_{\beta }\Phi \nabla^{\rho }\Phi \nn\\&&+ 8 e^{4 \Phi} R_{\kappa \lambda \tau \omega } F_{\gamma }{}^{\tau } F_{\mu \nu } F_{\rho }{}^{\omega } \nabla_{\beta }\Phi \nabla^{\rho }\Phi + 6 e^{6 \Phi} F_{\gamma \rho } F_{\kappa \lambda } F_{\mu }{}^{\tau } F_{\nu }{}^{\omega } F_{\tau \omega } \nabla_{\beta }\Phi \nabla^{\rho }\Phi \nn\\&&- 4 e^{6 \Phi} F_{\gamma }{}^{\tau } F_{\kappa \lambda } F_{\mu \nu } F_{\rho }{}^{\omega } F_{\tau \omega } \nabla_{\beta }\Phi \nabla^{\rho }\Phi + 8 e^{4 \Phi} R_{\kappa \lambda \rho \omega } F_{\gamma }{}^{\tau } F_{\mu \nu } F_{\tau }{}^{\omega } \nabla_{\beta }\Phi \nabla^{\rho }\Phi\nn\\&& - 4 e^{4 \Phi} R_{\kappa \lambda \tau \omega } F_{\gamma \rho } F_{\mu \nu } F^{\tau \omega } \nabla_{\beta }\Phi \nabla^{\rho }\Phi + e^{6 \Phi} F_{\gamma \rho } F_{\kappa \lambda } F_{\mu \nu } F_{\tau \omega } F^{\tau \omega } \nabla_{\beta }\Phi \nabla^{\rho }\Phi \nn\\&&- 4 e^{2 \Phi} R_{\kappa \lambda \tau \omega } R_{\mu \nu }{}^{\tau \omega } \nabla_{\gamma }F_{\beta \rho } \nabla^{\rho }\Phi - 2 e^{6 \Phi} F_{\beta }{}^{\tau } F_{\gamma }{}^{\omega } F_{\mu \nu } F_{\tau \omega } \nabla_{\lambda }F_{\kappa \rho } \nabla^{\rho }\Phi \nn\\&&-  e^{6 \Phi} F_{\beta \gamma } F_{\mu \nu } F_{\tau \omega } F^{\tau \omega } \nabla_{\lambda }F_{\kappa \rho } \nabla^{\rho }\Phi - 4 e^{6 \Phi} F_{\beta }{}^{\tau } F_{\gamma }{}^{\omega } F_{\mu \nu } F_{\rho \tau } \nabla_{\lambda }F_{\kappa \omega } \nabla^{\rho }\Phi \nn\\&&- 4 e^{6 \Phi} F_{\beta \rho } F_{\gamma }{}^{\tau } F_{\mu \nu } F_{\tau }{}^{\omega } \nabla_{\lambda }F_{\kappa \omega } \nabla^{\rho }\Phi - 4 e^{6 \Phi} F_{\beta \gamma } F_{\mu \nu } F_{\rho }{}^{\tau } F_{\tau }{}^{\omega } \nabla_{\lambda }F_{\kappa \omega } \nabla^{\rho }\Phi \nn\\&&+ 4 e^{4 \Phi} R_{\kappa \lambda \tau \omega } F_{\beta }{}^{\tau } F_{\gamma }{}^{\omega } \nabla_{\nu }F_{\mu \rho } \nabla^{\rho }\Phi + 4 e^{4 \Phi} R_{\kappa \lambda \tau \omega } F_{\beta \gamma } F^{\tau \omega } \nabla_{\nu }F_{\mu \rho } \nabla^{\rho }\Phi \nn\\&&- 16 e^{4 \Phi} F_{\gamma }{}^{\tau } \nabla_{\beta }\Phi \nabla_{\lambda }F_{\kappa \tau } \nabla_{\nu }F_{\mu \rho } \nabla^{\rho }\Phi - 8 e^{4 \Phi} R_{\kappa \lambda \rho \omega } F_{\beta }{}^{\tau } F_{\gamma }{}^{\omega } \nabla_{\nu }F_{\mu \tau } \nabla^{\rho }\Phi \nn\\&&- 8 e^{4 \Phi} R_{\kappa \lambda \rho \omega } F_{\beta \gamma } F^{\tau \omega } \nabla_{\nu }F_{\mu \tau } \nabla^{\rho }\Phi - 2 e^{2 \Phi} R_{\kappa \lambda \tau \omega } R_{\mu \nu }{}^{\tau \omega } F_{\beta \gamma } \nabla_{\rho }\Phi \nabla^{\rho }\Phi \nn\\&&+ 2 e^{4 \Phi} R_{\kappa \lambda \tau \omega } F_{\beta }{}^{\tau } F_{\gamma }{}^{\omega } F_{\mu \nu } \nabla_{\rho }\Phi \nabla^{\rho }\Phi -  e^{6 \Phi} F_{\beta }{}^{\tau } F_{\gamma }{}^{\omega } F_{\kappa \lambda } F_{\mu \nu } F_{\tau \omega } \nabla_{\rho }\Phi \nabla^{\rho }\Phi \nn\\&&+ 2 e^{4 \Phi} R_{\kappa \lambda \tau \omega } F_{\beta \gamma } F_{\mu \nu } F^{\tau \omega } \nabla_{\rho }\Phi \nabla^{\rho }\Phi -  \frac{1}{2} e^{6 \Phi} F_{\beta \gamma } F_{\kappa \lambda } F_{\mu \nu } F_{\tau \omega } F^{\tau \omega } \nabla_{\rho }\Phi \nabla^{\rho }\Phi \nn\\&&- 8 e^{4 \Phi} F_{\gamma }{}^{\tau } F_{\mu \nu } \nabla_{\beta }\Phi \nabla_{\lambda }F_{\kappa \tau } \nabla_{\rho }\Phi \nabla^{\rho }\Phi + 8 e^{2 \Phi} R_{\kappa \lambda \tau \omega } R_{\mu \nu \rho }{}^{\omega } \nabla^{\rho }\Phi \nabla^{\tau }F_{\beta \gamma } \nn\\&&+ 2 e^{4 \Phi} R_{\kappa \lambda \rho \omega } F_{\beta }{}^{\rho } \nabla_{\nu }F_{\tau }{}^{\omega } \nabla^{\tau }F_{\gamma \mu } + 4 e^{4 \Phi} R_{\kappa \lambda \tau \omega } F_{\beta }{}^{\rho } \nabla_{\gamma }F_{\rho }{}^{\omega } \nabla^{\tau }F_{\mu \nu } \nn\\&&- 4 e^{4 \Phi} F_{\gamma \rho } \nabla_{\beta }\Phi \nabla_{\lambda }F_{\kappa \tau } \nabla^{\rho }\Phi \nabla^{\tau }F_{\mu \nu } + 4 e^{4 \Phi} \nabla_{\gamma }F_{\beta \rho } \nabla_{\lambda }F_{\kappa \tau } \nabla^{\rho }\Phi \nabla^{\tau }F_{\mu \nu } \nn\\&&+ 2 e^{4 \Phi} F_{\beta \gamma } \nabla_{\lambda }F_{\kappa \tau } \nabla_{\rho }\Phi \nabla^{\rho }\Phi \nabla^{\tau }F_{\mu \nu } + 4 e^{4 \Phi} R_{\kappa \lambda \rho \omega } F_{\beta \gamma } \nabla_{\nu }F_{\tau }{}^{\omega } \nabla^{\tau }F_{\mu }{}^{\rho } \nn\\&&+ 8 e^{2 \Phi} R_{\kappa \lambda \tau \omega } R_{\mu \nu \rho }{}^{\omega } F_{\beta \gamma } \nabla^{\rho }\Phi \nabla^{\tau }\Phi - 8 e^{4 \Phi} R_{\kappa \lambda \tau \omega } F_{\beta \rho } F_{\gamma }{}^{\omega } F_{\mu \nu } \nabla^{\rho }\Phi \nabla^{\tau }\Phi \nn\\&&- 8 e^{4 \Phi} R_{\kappa \lambda \tau \omega } F_{\beta \gamma } F_{\mu \nu } F_{\rho }{}^{\omega } \nabla^{\rho }\Phi \nabla^{\tau }\Phi + 4 e^{6 \Phi} F_{\beta \rho } F_{\gamma }{}^{\omega } F_{\kappa \lambda } F_{\mu \nu } F_{\tau \omega } \nabla^{\rho }\Phi \nabla^{\tau }\Phi \nn\\&&+ 2 e^{6 \Phi} F_{\beta \gamma } F_{\kappa \lambda } F_{\mu \nu } F_{\rho }{}^{\omega } F_{\tau \omega } \nabla^{\rho }\Phi \nabla^{\tau }\Phi - 16 e^{4 \Phi} F_{\gamma \rho } F_{\mu \nu } \nabla_{\beta }\Phi \nabla_{\lambda }F_{\kappa \tau } \nabla^{\rho }\Phi \nabla^{\tau }\Phi\nn\\&& + 8 e^{4 \Phi} F_{\beta \gamma } \nabla_{\lambda }F_{\kappa \tau } \nabla_{\nu }F_{\mu \rho } \nabla^{\rho }\Phi \nabla^{\tau }\Phi - 8 e^{4 \Phi} F_{\gamma \tau } F_{\kappa \lambda } F_{\mu \nu } \nabla_{\beta }\Phi \nabla_{\rho }\Phi \nabla^{\rho }\Phi \nabla^{\tau }\Phi \nn\\&&+ 8 e^{4 \Phi} F_{\beta \gamma } F_{\mu \nu } \nabla_{\lambda }F_{\kappa \tau } \nabla_{\rho }\Phi \nabla^{\rho }\Phi \nabla^{\tau }\Phi + 2 e^{4 \Phi} F_{\beta \gamma } F_{\kappa \lambda } F_{\mu \nu } \nabla_{\rho }\Phi \nabla^{\rho }\Phi \nabla_{\tau }\Phi \nabla^{\tau }\Phi \nn\\&&+ 2 e^{6 \Phi} F_{\beta }{}^{\rho } F_{\gamma }{}^{\tau } F_{\rho }{}^{\omega } \nabla_{\nu }F_{\mu \tau } \nabla_{\omega }F_{\kappa \lambda } + \frac{1}{2} e^{4 \Phi} R_{\kappa \lambda \rho \tau } F^{\rho \tau } \nabla_{\nu }F_{\mu \omega } \nabla^{\omega }F_{\beta \gamma } \nn\\&&+ 2 e^{6 \Phi} F_{\beta }{}^{\rho } F_{\gamma \mu } F_{\rho }{}^{\tau } \nabla_{\nu }F_{\tau \omega } \nabla^{\omega }F_{\kappa \lambda } + 16 e^{4 \Phi} R_{\kappa \lambda \tau \omega } F_{\gamma }{}^{\rho } F_{\rho }{}^{\tau } \nabla_{\beta }\Phi \nabla^{\omega }F_{\mu \nu } \nn\\&&- 2 e^{4 \Phi} R_{\kappa \lambda \tau \omega } F^{\rho \tau } \nabla_{\gamma }F_{\beta \rho } \nabla^{\omega }F_{\mu \nu } + 4 e^{4 \Phi} R_{\kappa \lambda \tau \omega } F_{\beta \rho } F_{\gamma }{}^{\tau } \nabla^{\rho }\Phi \nabla^{\omega }F_{\mu \nu } \nn\\&&+ 4 e^{4 \Phi} R_{\kappa \lambda \tau \omega } F_{\beta \gamma } F_{\rho }{}^{\tau } \nabla^{\rho }\Phi \nabla^{\omega }F_{\mu \nu } -  \frac{1}{4} e^{6 \Phi} F_{\beta }{}^{\rho } F_{\gamma }{}^{\tau } F_{\rho \tau } \nabla_{\omega }F_{\kappa \lambda } \nabla^{\omega }F_{\mu \nu }\nn\\&& -  \frac{1}{8} e^{6 \Phi} F_{\beta \gamma } F_{\rho \tau } F^{\rho \tau } \nabla_{\omega }F_{\kappa \lambda } \nabla^{\omega }F_{\mu \nu } + 4 e^{4 \Phi} R_{\kappa \lambda \tau \omega } F_{\beta \gamma } \nabla^{\tau }F_{\mu }{}^{\rho } \nabla^{\omega }F_{\nu \rho }\nn\\&& - 4 e^{4 \Phi} R_{\kappa \lambda \rho \omega } F_{\beta \gamma } \nabla^{\tau }F_{\mu }{}^{\rho } \nabla^{\omega }F_{\nu \tau } + 2 e^{4 \Phi} R_{\kappa \lambda \tau \omega } F_{\beta }{}^{\rho } \nabla_{\mu }F_{\gamma \rho } \nabla^{\omega }F_{\nu }{}^{\tau }.
\eeqa
The action in equation \reef{reduce1} is, in fact, equivalent to the action in \reef{reduce}, up to non-gauge-invariant total derivative terms. While the first term in \reef{reduce1} represents the standard Chern-Simons couplings in type IIA theory—previously found in \cite{Duff:1995wd} from M2-brane/M5-brane duality—the remaining terms are new couplings that have not appeared in the literature. These couplings all vanish in the absence of RR field strength $F_{\mu\nu}$. Note that there is an even number of RR fields in \reef{reduce1}. The Levi-Civita tensor and the \(B\)-field have even parity, while the RR field strengths \(F^{(1)}\) and \(\bar{F}^{(4)}\) have odd parity. Hence, the couplings in \reef{reduce1} are invariant under parity. This is expected because the original 11-dimensional Chern-Simons coupling is also invariant under parity.

Note that all indices of the RR field strength $\bar{F}^{(4)}$ and the NS-NS field strength $H^{(3)}$ in \reef{reduce1} are contracted with the Levi-Civita tensor. One may speculate that other couplings could exist where these field strengths are also contracted with other tensors. Such terms might originate from other gauge-invariant couplings in M-theory where the three-form field strength $dA^{(3)}$ appears linearly in the effective action.
We have verified that if one considers all possible contractions of a single 11-dimensional Levi-Civita tensor with $dA^{(3)}$ and the Riemann curvature tensor at eighth derivative order, then, after applying the 11-dimensional Bianchi identities, no such coupling survives. Consequently, the couplings in equation \reef{reduce1} represent the complete set of one-loop couplings in type IIA theory that include the RR field strength $\bar{F}^{(4)}$ and the NS-NS field strength $H^{(3)}$ linearly.

It is also worth noting a key difference in the dimensional reduction of the 11-dimensional couplings based on the number of the three-form $A^{(3)}$: couplings with an even number of $A^{(3)}$ can produce direct, analogous couplings in ten dimensions—as well as many others—particularly in cases where no tensor carries the Killing index $y$ (see \cite{Garousi:2025wfk} for the reduction of the pure gravity part); in contrast, couplings with an odd number of $A^{(3)}$, which must be accompanied by the 11-dimensional Levi-Civita symbol, cannot produce direct analogues of the original eleven-dimensional coupling. In fact, none of the resulting ten-dimensional couplings is an analog of the original, since the Levi-Civita symbol always contributes the Killing index $y$ during reduction. This index is subsequently removed, transforming the symbol into its ten-dimensional counterpart and altering the structure of the coupling. For instance, in \reef{reduce1}, there is no coupling involving one three-form and four Riemann curvatures, whereas such a coupling exists in the 11-dimensional theory.

\section{Reduction of type IIA on K3 and S-duality in 6D}

Type IIA string theory compactified on a K3 surface is known to be dual to heterotic string theory on $T^4$ (see, e.g., \cite{Becker:2007zj}). Each theory contains 81 scalars; however, we are interested in only one specific scalar from each—the dilaton. Similarly, while there are 24 vector fields in each theory, we focus on two vectors from each. In type IIA theory, these are the RR one-form and the gauge field arising from the Hodge dualization of the RR four-form. In the heterotic theory, the relevant vectors are one from a component of the metric along one circle of the $T^4$, and another from a component of the $B$-field along the same circle. The field content of both theories also includes a metric and a Kalb-Ramond field ($H$-field).

For this duality to hold, it must map the specific fields of the six-dimensional type IIA theory to their heterotic counterparts, ensuring their effective actions match at every derivative order. The dilaton's sign reverses under this map \cite{Liu:2013dna}, identifying the duality as S-duality. Because the dilaton sets the coupling constant, this sign reversal means the duality connects a weakly coupled theory to a strongly coupled one. Therefore, studying the effective action under S-duality requires including all loop and non-perturbative corrections at each derivative order before the symmetry can be imposed. An exception occurs when a coupling has an exact dilaton dependence, which allows for its direct study under S-duality \cite{Tseytlin:1995fy}.

The Chern-Simons coupling at order $\ell_p^6$ in M-theory is unique; hence, the corresponding one-loop coupling in type IIA theory (see \ref{reduce1}) is exact. The K3 reduction of these couplings produces four-derivative and higher-order couplings in six dimensions. The four-derivative couplings are expected to be exact (see, e.g., \cite{Liu:2019ses}). On the other hand, the heterotic theory has tree-level couplings at four-derivative order in the NS-NS sector that receive no loop corrections \cite{Ellis:1987dc,Ellis:1989fi}. The $T^4$ reduction of these couplings is also exact. Therefore, the four-derivative couplings in the type IIA theory should map to the four-derivative couplings in the heterotic theory under the duality.

In the next subsection, we find the six-dimensional four-derivative couplings corresponding to the type IIA couplings in \reef{reduce1} and determine their S-duality transformations. In Subsection 3.2, we consider the reduction of the heterotic coupling $H_{\mu\nu\alpha}\Omega^{\mu\nu\alpha}$ on $T^4$ and compare it with the S-dual of the type IIA couplings.

\subsection{Type IIA reduction on K3}

The K3 reduction of the eight-derivative couplings in \reef{reduce1} generates both the four-derivative couplings in which we are interested and higher-derivative couplings in which we are not. To isolate the four-derivative couplings, we consider an ansatz where the metric takes the block-diagonal form:
\beqa
ds^2 = G^6_{\mu\nu}(x)dx^\mu dx^\nu + g^4_{ab}(y)dy^a dy^b,\label{K3red}
\eeqa
with $y^a$ denoting the K3 coordinates. In this section, we use the indices $\mu,\nu,\cdots$ for the 10-dimensional and 6-dimensional spaces and the indices $a,b,\cdots$ for the compact 4-dimensional space. For the block-diagonal metric, the 10-dimensional Levi-Civita symbol can be written as the product of the 6-dimensional and 4-dimensional Levi-Civita symbols. In terms of the Levi-Civita tensor, this is expressed as:
\beqa
\sqrt{-G}\,\epsilon_{10}&=&\sqrt{-G^6}\,\epsilon_{6}\ \sqrt{g^4}\,\epsilon_{4}.
\eeqa
The non-flatness of the K3 surface introduces non-vanishing curvature contributions. In particular, the integral of the first Pontryagin class over K3 is (see \eg \cite{Liu:2019ses}):
\beqa
\frac{1}{32\pi^2}\int_{K3} d^4y\sqrt{g^4}\,\epsilon_4^{abcd}R_{abef}R_{cd}{}^{ef}&=&48.\labell{K3int}
\eeqa
This topological constraint plays a key role in producing the four-derivative couplings when applied to the eight-derivative couplings in \reef{reduce1}.

Using the above constraint, the K3 reduction of the ten-dimensional, one-loop gravity couplings in \reef{reduce1} produces the following four-derivative term in six dimensions:
\beqa
S_{6D}&\!\!\!\!=\!\!\!\!&-\frac{2}{\kappa^2}\frac{\pi^4\ell_s^6}{12}\int d^{6}x\sqrt{-G^6}\,\epsilon_6^{\alpha\beta\gamma\mu\delta\sigma}\Big[6 B_{\alpha \beta } R_{\gamma \mu }{}^{\rho \tau } R_{\delta \sigma \rho \tau } + \frac{1}{2} e^{4 \Phi} 
\bar{F}_{\gamma \mu \delta \sigma } F_{\alpha }{}^{\rho } F_{\beta 
}{}^{\tau } F_{\rho \tau }\nn\\&& -  e^{2 \Phi} \bar{F}_{\gamma \mu 
\delta \sigma } R_{\alpha \beta \rho \tau } F^{\rho 
\tau } + \frac{1}{4} e^{4 \Phi} \bar{F}_{\gamma \mu \delta \sigma 
} F_{\alpha \beta } F_{\rho \tau } F^{\rho \tau } - 4 e^{2 
\Phi} H_{\mu \delta \sigma } F_{\alpha }{}^{\rho } 
\nabla_{\gamma }F_{\beta \rho }\nn\\&& - 4 e^{2 \Phi} \bar{F}_{\gamma \mu 
\delta \sigma } F_{\beta \rho } \nabla_{\alpha }\Phi \nabla^{
\rho }\Phi + 4 e^{2 \Phi} \bar{F}_{\gamma \mu \delta \sigma } 
\nabla_{\beta }F_{\alpha \rho } \nabla^{\rho }\Phi + 2 e^{2 
\Phi} \bar{F}_{\gamma \mu \delta \sigma } F_{\alpha \beta } 
\nabla_{\rho }\Phi \nabla^{\rho }\Phi \Big].\labell{SR2}
\eeqa
To convert the first term into the one that appears in the heterotic theory, we use the following identity:
\beqa
\frac{1}{4}\epsilon_6^{\alpha\beta\gamma\mu\delta\sigma}R_{\kappa\lambda\alpha\beta}R^{\kappa\lambda}{}_{\gamma\mu}&=&\epsilon_6^{\alpha\beta\gamma\mu\delta\sigma}\nabla_{\mu}\Omega_{\alpha\beta\gamma},
\eeqa
where $\Omega_{\mu\nu\alpha}$ is the Lorentz Chern-Simons three-form in six dimensions, defined as:
\beqa \Omega_{\mu\nu\alpha} &=& \omega_{[\mu {\mu_1}}{}^{\nu_1}\partial_\nu \omega_{\alpha] {\nu_1}}{}^{\mu_1} + \frac{2}{3} \omega_{[\mu {\mu_1}}{}^{\nu_1} \omega_{\nu {\nu_1}}{}^{\alpha_1} \omega_{\alpha]{\alpha_1}}{}^{\mu_1} \,\,;\,\,\, \omega_{\mu {\mu_1}}{}^{\nu_1} = e^\nu{}_{\mu_1} \nabla_\mu e_\nu{}^{\nu_1},\labell{Omeg}
 \eeqa 
where $\mu_1,\nu_1,\cdots$ represent indices in the flat tangent space, and $e_\mu{}^{\mu_1} e_\nu{}^{\nu_1} \eta_{\mu_1\nu_1} = G^6_{\mu\nu}$.
Then the first term in \reef{SR2} can be written in terms of $\epsilon_6^{\alpha\beta\gamma\mu\delta\sigma}H_{\alpha\beta\gamma}\Omega_{\mu\delta\sigma}$ up to a total derivative term.

We now transform the six-dimensional one-loop theory \reef{SR2} under S-duality. It has been observed in \cite{Garousi:2025xqn} that S-duality in type I/heterotic theory requires $\Omega$ to be invariant. We expect the same to hold for the six-dimensional case. The S-duality transformations are as follows:
\beqa
&&H^{\mu\nu\alpha}=\frac{e^{-2\Phi'}}{3!}\epsilon_6^{\mu\nu\alpha\beta\gamma\lambda}\bar{H}_{\beta\gamma\lambda}\,,\,\bar{F}^{\mu\nu\alpha\beta}=\frac{1}{2}W_{\gamma\lambda}\epsilon_6^{\gamma\lambda\mu\nu\alpha\beta}\,,\,F_{\alpha\beta}=V_{\alpha\beta}\,,\nn\\&&G_{\alpha\beta}=e^{-2\Phi'}G'_{\alpha\beta}\,,\,\Phi=-\Phi'\,,\,\Omega_{\alpha\beta\gamma}\,=\,\bar{\Omega}_{\alpha\beta\gamma},
\eeqa
where $\bar{H}, W, V, \Phi',G',\bar{\Omega}$ are fields in the dual theory. Under the above transformation, one finds the six-dimensional action \reef{SR2} transforms to the following dual action:
\beqa
S^{dual}_{6D}&=&-\frac{2}{\kappa^2}\frac{\pi^4\ell_s^6}{12}\int d^{6}x\sqrt{-G'}\,e^{-2\Phi'}\Big[  48 \bar{H}^{\alpha \beta \gamma } \bar{\Omega} 
_{\alpha \beta \gamma }-12 V_{\alpha }{}^{\gamma } V^{\alpha \beta } V_{\beta 
}{}^{\mu } W_{\gamma \mu } \nn\\&&- 6 V_{\alpha \beta } V^{\alpha 
\beta } V^{\gamma \mu } W_{\gamma \mu } + 24 
R'_{\alpha \beta \gamma \mu } V^{\alpha \beta } 
W^{\gamma \mu } + 24 
\bar{H}_{\beta \gamma \mu } V^{\alpha \beta } \nabla^{\mu 
}V_{\alpha }{}^{\gamma } \nn\\&&- 48 \bar{H}_{\beta \gamma \mu } V_{\alpha 
}{}^{\beta } V^{\gamma \mu } \nabla^{\alpha }\Phi' - 96 
W^{\beta \gamma } \nabla^{\alpha }\Phi' \nabla_{\gamma 
}V_{\alpha \beta } + 96 V^{\alpha \beta } W_{\alpha 
}{}^{\gamma } \nabla_{\gamma }\nabla_{\beta }\Phi' \Big].\labell{SR21}
\eeqa
Note that the overall dilaton factor $e^{-2\Phi'}$ indicates the dual action is at the sphere level. Note also that the above action is linear in the field strengths $\bar{H}$ and $W$, so it should correspond to couplings in the heterotic theory that are linear in the NS-NS antisymmetric tensor field strength. In the next subsection, we consider such couplings in the heterotic theory.

\subsection{Heterotic reduction on $T^4$}

The heterotic string theory features 496 massless vector fields in the adjoint representation of the $SO(32)$ or $E_8\times E_8$ gauge group, as well as NS-NS fields that are singlets under these groups. For the purposes of this paper, we consider the ten-dimensional vector gauge fields to be zero. In this case, the nonstandard local Lorentz transformation for the $B$-field requires a specific field strength in the effective action, as described in \cite{Green:1984sg}:
\beqa
\hat{H}_{\mu\nu\alpha}&=& H'_{\mu\nu\alpha}-\frac{3}{2}\alpha' \Omega'_{\mu\nu\alpha}\,.
\labell{replace}
\eeqa
Here, $H'_{\mu\nu\alpha}=3\partial_{[\mu}B'_{\nu\alpha]}$, and $\Omega'$ is the ten-dimensional Chern-Simons three-form defined in \reef{Omeg}.

The leading-order effective action for the heterotic theory, with Yang-Mills fields set to zero, is
\beqa
\bS'=-\frac{2}{\kappa'^2_{10}}\int d^{10}x \sqrt{-G_{10}'} e^{-2\Phi'}\! \left( R' + 4\nabla_{\mu}\Phi' \nabla^{\mu}\Phi'-\frac{1}{12}\hat{H}^2\right) ,\labell{baction}
\eeqa
where the ten-dimensional gravitational coupling is given by $\kappa'^2_{10}=\frac{1}{\pi}(2\pi\ell'_s)^8g'_s{}^2$, with $\ell_s'$ being the heterotic string length scale. This action produces the following four-derivative coupling that is linear in the $H'$-field:
\beqa
\bS'_{H\Omega}&=&-\frac{2}{\kappa'^2_{10}}\frac{\ell'^2_s}{8}\int d^{10}x\sqrt{-G_{10}'}e^{-2\Phi'}\Big[2H'^{\alpha \beta \gamma } \Omega' 
_{\alpha \beta \gamma }\Big]\,.\labell{SHet}
\eeqa 
The circular reduction of this classical coupling and its symmetry under T-duality have been studied in \cite{Garousi:2023pah}.

To study the KK reduction of this coupling on $T^4$ for zero moduli and with only two KK vectors—resulting from the metric and the $B$-field along a single circle—we decompose \( T^4 = S^1 \times T^3 \) , where $T^3$ has a constant diagonal metric, and write the 10-dimensional metric as follows:
\beqa
ds'^2 = G'_{MN}(x)dx^M dx^N + g'_{ij}(z)dz^i dz^j,\labell{dsHet}
\eeqa
where $z^i$ are the coordinates of $T^3$, and $x^M$ denotes both the circle coordinate of $S^{(1)}$ and the six-dimensional spacetime coordinates. Performing the KK reduction on the seven-dimensional metric using the ansatz:
\beqa  
G'_{MN} = \left(\matrix{G'_{\mu\nu} + R_1^{-2}g_{\mu} g_{\nu} &   g_{\mu} \cr  g_{\nu} & R_1^2 &}\!\!\!\!\!\right),  
  \labell{reduc1}  
\eeqa  
where $R_1$ represents the radius of the circle coordinate $y$, while indices $\mu,\nu$ label the six-dimensional spacetime directions orthogonal to $y$. The dimensional reduction then yields the following effective action in six dimensions:
\beqa
S'_{H\Omega}&=&-\frac{2}{\kappa'^2_{10}}\frac{\ell'^2_sV'}{8}\int d^{6}x\sqrt{-G'}e^{-2\Phi'}\Big[- \frac{1}{2} V_{\alpha }{}^{\gamma } V^{\alpha \beta } 
V_{\beta }{}^{\mu } W_{\gamma \mu } -  \frac{1}{4} V_{\alpha 
\beta } V^{\alpha \beta } V^{\gamma \mu } W_{\gamma \mu }\nn\\&& + R_{\alpha \gamma \beta \mu } V^{\alpha \beta } 
W^{\gamma \mu } + 2 \bar{H}^{\alpha \beta \gamma } \bar{\Omega 
}_{\alpha \beta \gamma } + \frac{1}{2} \bar{H}_{\gamma \mu \sigma} 
V^{\alpha \beta } V^{\gamma \mu }\bar{\omega}^{\sigma}{}_{\alpha 
\beta } + V^{\alpha \beta } W^{\gamma \mu }\bar{\omega}_{\gamma 
\alpha }{}^{\sigma}\bar{\omega}_{\mu \beta \sigma}\nn\\&& - 2 W^{\alpha \beta } 
\bar{\omega}_{\alpha }{}^{\gamma \mu } \nabla_{\mu }V_{\beta 
\gamma } + \bar{H}_{\beta \gamma \mu } V^{\alpha \beta } 
\nabla^{\mu }V_{\alpha }{}^{\gamma }\Big]\,,\labell{S'6}
\eeqa 
where $\bar{\omega}_{\mu\nu\alpha}$ is related to the six-dimensional spin connection as $\bar{\omega}_{\mu\nu\alpha}=\bar{\omega}_{\mu\mu_1\nu_1}e^{\mu_1}_\nu e^{\nu_1}_\alpha$. In the above equation, $V'=\int_{S^{(1)}} R_1dy \int_{T^3} d^3 z \sqrt{g'}$ represents the $T^4$ volume. This reduction is identical to the circular reduction found in \cite{Garousi:2023pah} for a fixed radius. There are terms in the above reduction that involve the spin connection. These terms are not invariant under local Lorentz transformations and should be removed by field redefinitions that include the spin connection.

To study the field redefinition, we must consider the reduction of the leading-order, two-derivative terms to six dimensions. The KK reduction of the two-derivative terms in \reef{baction} for a fixed circle radius uses the metric reduction in \reef{reduc1} and the following $B$-field reduction ansatz from \cite{Kaloper:1997ux}:
\beqa
B_{MN}= \left(\matrix{\bb_{\alpha\beta}+\frac{1}{2}b_{\alpha }g_{\beta }- \frac{1}{2}b_{\beta }g_{\alpha }&b_{\alpha }\cr - b_{\beta }&0&}\right).\labell{Breduce}
\eeqa
This reduction produces the following two-derivative action in six dimensions:
 \beqa
{ S'} = -\frac{2V'}{\kappa'^2_{10}}\int d^6 x \sqrt{-G'} e^{-2\Phi'} \left[ R' + 4\nabla_\mu \Phi' \nabla^\mu \Phi' - \frac{1}{4} W_{\mu\nu} W^{\mu\nu}- \frac{1}{4} V_{\mu\nu} V^{\mu\nu}-\frac{1}{12}\bar{H}_{\mu\nu\alpha} \bar{H}^{\mu\nu\alpha}\right]\!, \label{Het}
\eeqa
where $W_{\mu\nu}=\prt_\mu b_\nu-\prt_\nu b_\mu$, $V_{\mu\nu}=\prt_\mu g_\nu-\prt_\nu g_\mu$, and the three-form $\bH$ is defined as $\bH_{\mu\nu\alpha}=\tilde{H}_{\mu\nu\alpha}-g_{\mu}W_{\nu\alpha}-g_{\alpha}W_{\mu\nu}-g_{\nu}W_{\alpha\mu}$. Here, the three-form $\tilde{H}$ is the field strength of the two-form $\bb_{\mu\nu}+\frac{1}{2}b_{\mu}g_{\nu}-\frac{1}{2}b_\nu g_\mu$ in \reef{Breduce}. The three-form $\bH$ is not the field strength of a two-form and satisfies the following Bianchi identity \cite{Kaloper:1997ux}:
\beqa
\prt_{[\mu} \bH_{\nu\alpha\beta]}&=&-\frac{3}{2}V_{[\mu\nu}W_{\alpha\beta]}\,.\labell{anB}
\eeqa
The field redefinition is constrained by the requirement that it must satisfy the above Bianchi identity.

In general, one may consider field redefinitions for all six-dimensional fields. However, for the purposes of this paper, we only need to consider the field redefinition for the vector field $b_\alpha$, i.e.,
\beqa
b_\alpha=b_\alpha+\ell'^2_s\Delta b_\alpha,\labell{bfr}
\eeqa
where $\Delta b_\alpha$ consists of fields at the two-derivative order. The Bianchi identity \reef{anB} indicates that the field redefinition of $\bH$ is related to this field redefinition by $\Delta\bH_{\mu\nu\alpha}=-3V_{[\mu\nu}\Delta b_{\alpha]}$. This redefinition then produces the following couplings at the four-derivative order:
\beqa
\Delta S'&=&-\frac{2\ell'^2_sV'}{\kappa'^2_{10}}\int d^{6}x\sqrt{-G'}e^{-2\Phi'}\Big[\frac{1}{2} \bar{H}_{\alpha\beta\gamma} V^{\beta\gamma} \Delta b^{\alpha} + W_{\alpha\beta} \nabla^{\beta}(\Delta b^{\alpha})\Big].\labell{DS0}
\eeqa
The action \reef{S'6} for the field in \reef{bfr} then includes the four-derivative couplings $S'_{H\Omega}+\Delta S'$.

To remove the terms in \reef{S'6} that involve the spin connection, we consider the following field redefinition:
\beqa
\Delta b_\alpha&=&-\bar{\omega}_{\alpha\beta\gamma}V^{\beta\gamma}.
\eeqa
The corresponding  $\Delta S'$ is
\beqa
\Delta S'&\!\!\!\!=\!\!\!\!&-\frac{2\ell'^2_sV'}{\kappa'^2_{10}}\int d^{6}x\sqrt{-G'}e^{-2\Phi'}\Big[- \frac{1}{2} \bar{H}_{\gamma \mu \nu } V^{\alpha \beta } 
V^{\gamma \mu }\bar{\omega}^{\nu }{}_{\alpha \beta } -  
W^{\alpha \beta }\bar{\omega}_{\alpha }{}^{\gamma \mu } \nabla_{
\beta }V_{\gamma \mu } -  V^{\alpha \beta } W^{\gamma \mu } 
\nabla_{\mu }\bar{\omega }_{\gamma \alpha \beta }\Big].\nn
\eeqa
Adding this term to \reef{S'6} results in the removal of the spin connection terms. That is 
\beqa
S'_{H\Omega}+\Delta S'&=&-\frac{2}{\kappa'^2_{10}}\frac{\ell'^2_sV'}{8}\int d^{6}x\sqrt{-G'}e^{-2\Phi'}\Big[- \frac{1}{2} V_{\alpha }{}^{\gamma } V^{\alpha \beta } 
V_{\beta }{}^{\mu } W_{\gamma \mu } -  \frac{1}{4} V_{\alpha 
\beta } V^{\alpha \beta } V^{\gamma \mu } W_{\gamma \mu } \nn\\&&+ 2 
R_{\alpha \gamma \beta \mu } V^{\alpha \beta } 
W^{\gamma \mu } + 2 \bar{H}^{\alpha \beta \gamma } \bar{\Omega 
}_{\alpha \beta \gamma } -  \frac{1}{2} \bar{H}_{\beta \gamma \mu 
} V^{\alpha \beta } \nabla_{\alpha }V^{\gamma \mu }\Big]\,,\labell{S'61}
\eeqa 
where we have used the following relation between the Riemann curvature and the spin connection:
\beqa
R_{\alpha\beta\gamma\mu}&=&-\bar{\omega}_{\gamma \beta }{}^{\nu }\bar{\omega}_{\mu \alpha 
\nu } +\bar{\omega}_{\gamma \alpha }{}^{\nu }\bar{\omega}_{\mu 
\beta \nu } + \nabla_{\gamma }\bar{\omega}_{\mu \alpha \beta } - 
 \nabla_{\mu }\bar{\omega}_{\gamma \alpha \beta }.
\eeqa
As expected, the action in \reef{S'61} does not include the spin connection, other than that within the Chern-Simons three-form, and is consistent with local Lorentz gauge transformations. In fact, the $\bH\bar{\Omega}$ term can be combined with the $\bH^2$ term in the leading action to express $\bH$ in terms of the generalized field strength $\bH-3\alpha'\bar{\Omega}/2$, as in \reef{replace}.

The couplings in \reef{S'61} are now identical to those in the first two lines of \reef{SR21}, up to an overall factor. This overall factor also matches using the fact that the NS5-brane of heterotic theory, when wrapped on $T^4$, transforms under S-duality into the fundamental string of type IIA theory (see  \eg \cite{Becker:2007zj}). The equality of their tensions yields the relation:
\beqa
\frac{2\pi V'}{(2\pi\ell_s')^6g_s'^2}&=&\frac{1}{2\pi\ell_s^2}.\label{NS5F}
\eeqa
Using this relation, one finds that \reef{S'61} produces exactly the couplings in the first two lines of \reef{SR21}.

The couplings in the last line of \reef{SR21} are also reproduced by \reef{DS0} if one uses the following covariant field redefinition:
\beqa
\Delta b_\alpha&=&4V_{\alpha\beta}\nabla^\beta\Phi'.
\eeqa
Hence, our calculations confirm that the two actions \reef{SR2} and \reef{S'6} are indeed S-dual to each other when appropriate field redefinitions are included. This provides a nontrivial verification of S-duality between type IIA theory on K3 and heterotic theory on $T^4$. It also confirms the eight-derivative couplings in \reef{reduce1} that were found from the Chern-Simons term in M-theory at order $\ell_p^6$.

\section{Conclusion}

In this paper, we have employed circular reduction of the Chern-Simons term \( t_8\epsilon_{11} A R^4 \) in the effective action of M-theory at order \( \ell_p^6 \) to derive the corresponding one-loop couplings at order \( \alpha'^3 \) in type IIA string theory. These include the standard Chern-Simons term \( t_8\epsilon_{10} B R^4 \), along with 1,173 additional couplings expected to be gauge invariant up to total derivative terms. By incorporating appropriate non-gauge-invariant total derivative terms, we successfully rewrote the 1,173 couplings in terms of 91 gauge-invariant couplings, as presented in \reef{reduce1}. These couplings exhibit linear dependence on the NS-NS field strength \( H^{(3)} \) and the RR field strength \( \bar{F}^{(4)} \). Furthermore, explicit calculation verifies that no other couplings linear in the three-form \( A^{(3)} \) exist in M-theory at this order beyond the Chern-Simons term itself. This confirms that the new couplings we have found constitute the complete set of couplings in type IIA theory linear in \( H^{(3)} \) and \( \bar{F}^{(4)} \) at this order.

We also investigated the behavior of these new couplings under six-dimensional S-duality between type IIA theory on K3 and heterotic theory on \( T^4 \). Our results show that the four-derivative one-loop couplings in type IIA on K3 transform precisely under S-duality into the corresponding four-derivative tree-level couplings in heterotic theory on \( T^4 \).

It is well-established that the tree-level effective action of string theory is invariant under T-duality at each order in $\alpha'$ \cite{Sen:1991zi,Hohm:2014sxa}. Unlike S-duality, T-duality maps the tree-level action at a given order in $\alpha'$ to a tree-level action at the same order. A natural question is whether the one-loop effective action similarly transforms under T-duality into the one-loop action of the dual theory.
By examining the pure gravity couplings in type IIA theory at one-loop order, given by $(t_8t_8 - \frac{1}{8}\epsilon_{10}\epsilon_{10})R^4$, and their counterparts in type IIB theory, given by $(t_8t_8 + \frac{1}{8}\epsilon_{10}\epsilon_{10})R^4$, it appears unlikely that these satisfy T-duality in a straightforward manner \cite{Garousi:2024pqc}. To systematically investigate T-duality at the one-loop level, one could perform a circular reduction of the type IIA couplings in \reef{reduce1} to derive the corresponding nine-dimensional couplings. Applying T-duality to these results would then test whether the transformed couplings can be reproduced via circular reduction of gauge-invariant couplings in ten-dimensional type IIB theory.
While Chern-Simons-like terms are generally forbidden in type IIB theory \cite{Vafa:1995fj}, it has been argued that T-duality may necessitate their existence \cite{Antoniadis:1997eg}. Resolving this apparent contradiction represents a key challenge in understanding one-loop T-duality. We leave a detailed investigation of this calculation for future work.


\end{document}